\documentclass[]{spie}  

\usepackage{amsmath,amsfonts,amssymb}
\usepackage{graphicx}
\usepackage[colorlinks=true, allcolors=blue]{hyperref}

\title{Persistence characteristics of H4RG-15 detectors for ESO instruments}

\author[a]{Elizabeth M. George}
\author[a]{Domingo Alvarez Mendez}
\author[a]{Nagaraja Naidu Bezawada}
\author[a]{Derek J. Ives}
\author[a]{Mark Neeser}
\author[a]{Benoit Serra}
\affil[a]{European Southern Observatory, Karl-Schwarzschild-Str. 2, Garching bei M\"{u}nchen, Germany}

\authorinfo{Further author information: (Send correspondence to E.M.G.)\\E.M.G.: E-mail: egeorge@eso.org, Telephone: +49 89 3200 6347}

\begin{document} 
\maketitle

\begin{abstract}
As a part of the ESO Instrumentation Program, we already ordered 23 H4RG-15 (2.5 um cut-off) detectors for the instruments MOONS, MICADO, and HARMONI, with more of these detectors planned for future instruments on the ELT as well as possibly other ESO telescopes. As part of our comprehensive detector testing program, we characterize detector persistence at different temperatures between 40 and 85 K, allowing the best-informed choice of operating temperature for the science focal planes. We report the characteristics of persistence as a function of temperature for the 5 MOONS detectors and 10 MICADO detectors that have already been tested between 2019 and 2025. These detectors show a rich variety of persistence behaviour that can be fit with a single persistence model with varying parameters. We produce persistence trap maps and a model for persistence decay for each device, allowing automatic pipeline correction of persistence and also simulation of persistence behaviour of individual detectors using the Pyxel detector simulator. Each detector was unique and had different optimal operation temperatures for reduced persistence. We therefore recommend implementing a flexible thermal interface in the instrument design phase allowing for the detector operating temperature to be optimized after the science detectors have been fully characterized.
\end{abstract}


\section{INTRODUCTION}
\label{sec:intro}  
As a part of the ESO Instrumentation Program, we already ordered 23 H4RG-15 (2.5 um cut-off) detectors for the instruments MOONS, MICADO, and HARMONI, with more of these detectors planned for future instruments on the Extremely Large Telescopes (ELT) as well as possibly other ESO telescopes. All three of these instruments can be impacted by image persistence in their primary observation modes. Image persistence occurs when, during an exposure, some of the charge carriers are trapped within the depletion region, and detrap during subsequent exposures and are then recorded as signal.\cite{smith08} 

MOONS is a multi-object fiber-fed spectrograph on the VLT, covering the RI, YJ, and H bands at switchable resolutions ranging from approximately 4000 - 18000.\cite{cirasuolo16, cirasuolo20} In MOONS, there are four H4RG-15 detectors, each located in a separate camera with a standard operation temperature of 40 K. Two of these detectors cover the YJ-band and another 2 cover the H-band. Of particular importance for persistence is that the spectral format is not fixed in the H-band cameras, as the spectral resolution can be switched between high and low resolution, moving bright OH lines on the detector surface when operating modes are changed. Thus there is the risk of persistence from these bright lines contaminating regions normally free of OH lines in subsequent exposures.

MICADO is the first-light instrument for the ELT, which will operate in 4 modes: imaging, astrometry, coronography, and long-slit spectroscopy. The instrument is fed by the adaptive optics module MORFEO\cite{sturm24} The focal plane is comprised of nine H4RG-15 detectors in a close-butted 3x3 mosaic with a standard operating temperature of $82 \pm 2$K.\cite{bezawada23} A special read mode was developed with interleaved reset windows to decrease image persistence from user-defined known bright objects in imaging fields.\cite{bezawada26} However, persistence is still expected to impact observations, especially when switching between modes or after observing crowded fields where bright stars may cause persistence in subsequent exposures.

HARMONI is the first light, adaptive optics assisted, near-IR integral field spectrograph for the ELT. It covers a spectral range from 750 nm to 2450 nm with resolving powers from 3000 to 7000 and spatial sampling of 25 mas and 6 mas. It can operate in two adaptive optics modes - Single Conjugate Adaptive Optics (SCAO) including a high contrast capability as well as Multi-conjugate Adaptive Optics (MCAO). The project is resuming its final design phase after a rescope design phase in 2025.\cite{neichel26} There are eight H4RG-15 detectors arranged in four 2x1 focal plane mosaics.\cite{maciver26} Due to the several available diffraction gratings, from observation to observation the spectral format on the detectors will change, moving the physical location of the bright OH lines on the detector and creating image persistence in regions normally free of OH lines in subsequent exposures.

Given the potential of image persistence to impact science observations in all the instruments using H4RG-15 detectors, we have made strong efforts at ESO to characterize image persistence in all science detectors\cite{tulloch18}, develop a model to predict persistence impact\cite{tulloch19}, and develop a correction algorithm that can be applied to the science image pipeline to correct for any image persistence present in the data\cite{padovani22}. 

We are not the first, nor will we be the last to be affected by image persistence. The literature is full of papers describing efforts to characterize and model image persistence in HxRG detectors, including JWST,\cite{regan17, regan17a} Euclid,\cite{kubik24} Roman,\cite{mosby20, louie26} Hubble WFC3,\cite{long12, long13, long15} SALT RSS\cite{mosby16}, SPHEREX\cite{fazar25} and other H2RG detectors\cite{smith08a, baril08} as well as HdCdTe detectors outside the HxRG family (e.g. ALFA).\cite{legoff20,legoff21,legoff22} What we hope to gain from our approach is a method for characterizing and correcting for persistence that can be applied across ESO's large variety of instruments using HxRG detectors, which have been manufactured over many years in different batches.

In section \ref{sec:detectorsandtesting} we describe the detectors and test protocol used to characterize the detector persistence. Section \ref{sec:results} describes the different types of persistence behaviour seen in the 15 detectors tested to date. In section \ref{sec:modeling} we describe the model used to fit the data and discuss potential physical models that could explain the behaviour seen. Section \ref{sec:pipelincorrection} describes the pipeline correction algorithm developed to correct the science data and reports on the current use of that algorithm in ESO instruments. Finally in section \ref{sec:conclusion} we discuss our conclusions and future work.

\section{DETECTORS AND TESTING}
\label{sec:detectorsandtesting}

The first detectors delivered to ESO were the one engineering and four science grade H4RG-15 detectors for MOONS, delivered between 2019 and 2022. All detectors were characterized in the CRISLER test facility at ESO, a basic cryostat allowing images to be taken in the dark, with LED illumination, or through a pinhole with selectable J or H band filters, allowing approximate QE measurements using a blackbody. The first engineering detector was extensively tested between 2019 and 2021, allowing development of new modes of detector operation and resulting in some changes to the preamplifier electronics.\cite{ives20,alvarez23} 

The MOONS detectors were planned to be fully characterized at 40 and 80 K, with some parameters (such as persistence) tested at intermediate temperatures in some detectors. Performance characteristics of the MOONS detectors were generally excellent at 40 K despite the presence of increased crosshatch at 40 K seen at in a few detectors\cite{george24} and this was adopted as the baseline operating temperature for MOONS. Unfortunately due to tight AIT schedules, some detectors were only able to be tested at 40 K.

Nine science grade detectors for MICADO and one shared HARMONI/MICADO engineering grade detector were delivered to ESO between 2021 and 2025. These detectors were fully characterized at 40 and 85 K, with some parameters (such as persistence) tested also at intermediate temperatures. Bezawada et al. 2026\cite{bezawada26} gives an overview of all the performance characteristics tested, as well as describes a few features of the detectors seen. In particular, quantum efficiency measurements showed increased cross-hatch pattern at 40 K with a reduction in overall QE. Noise measurements also showed an increased contact resistance in some devices at 40 K, which has an impact on persistence in these devices, though in general noise performance improved as temperature decreased. 

MICADO is liquid nitrogen cooled, therefore the operating temperature of the detectors must be above 77K. The final operating temperature chosen was $82 \pm 2$K. HARMONI has a baseline detector operation temperature of 40 K, so all detectors were tested at this temperature as well to allow informed decisions in instrument design before the final HARMONI detectors became available for test.

Table \ref{tab:detectors} in appendix \ref{sec:appendixA} lists the detectors tested and the temperatures for which persistence measurements were obtained. We were particularly interested in temperatures around 40 K (baseline for HARMONI and MOONS) as well as temperatures around 80 K (baseline for MICADO), and decided to test 5K warmer as well (45K and 85 K) in case the thermal environment in the final instrument cryostat that would prevent operating at the lower temperature. Additionally, in the original testing of the first MOONS engineering device we saw an increase in persistence at intermediate temperatures, which we interpreted as a trap with energy around 0.13eV.\cite{ives20} We therefore adopted an intermediate test at 60K to look for evidence of this excess persistence, as not all detectors show this behaviour. 

\subsection{Persistence test algorithm}
\label{sec:algoritm}
We measure persistence in all detectors in two ways. We do a ``quick persistence verification'' test that involves illuminating the detectors with a bright flatfield image to accumulate approximately 50 ke- in 300s, then reset the detector and measure persistence in the next 300s dark image. This gives a standard test to compare detector-to-detector that can be performed quickly at different temperatures, and critically can be repeated by instrument teams after a detector is installed in the instrument and later at the telescope, allowing tracking of potential changes over time.\cite{bezawada26} This test is also used to verify persistence requirements from the instruments and to test the persistence correction algorithm. 

The subject of this paper is the result of a more involved persistence characterization that allows us to produce a trap map together with a spectrum of time constants. The development of this method followed an extensive investigation into the nature of charge trapping and detrapping in H2RG detectors conducted at ESO from 2016-2019.\cite{tulloch18, tulloch19} We refer readers interested in the details to these two papers and will not repeat the full findings here. From this work, we developed a standard test that allows us to produce a full trap map and spectrum of detrapping time constants for each detector. 

The persistence data is collected as follows: 
\begin{itemize}

    \item Let the detector sit in the dark for at least 1 day to fully detrap any charge.
    \item Reset the detector.
    \item Flash an LED to illuminate the detector up to the desired level (for a dark reference, no illumination; for a full measurement of all traps, up to full well).
    \item Leave the detector with charge on it for 10,000s (soak time).
    \item Reset the detector.
    \item Immediately begin reading out the detector non-destructively with increasing gaps between frames for approximately 14 hours. (e.g. first frames are read as quickly as possible to probe short time constant traps, frames in the middle of the sequence have gaps ranging from 10s of seconds to 10s of minutes, the last few frames are read an hour apart).
\end{itemize}

Each detector is tested at each temperature with no illumination and then again with full illumination. To analyze the data, we subtract the no illumination frames from the full illumination frames to remove the considerable effect of dark current over the 14 hour exposure. Because the characterization takes so long to run, this is usually only done once per temperature. This means that our signal-to-noise is generally not good enough to robustly fit the persistence signal in each pixel individually, as was done in LeGoff et al. 2020.\cite{legoff20} 

\subsection{Persistence model}
\label{sec:persistencemodel}
Following the methodology developed in Tulloch and George 2019,\cite{tulloch19} we co-add regions of between 300x300 and 550x550 pixels in regions of the detectors with similar persistence behaviour to get high signal-to-noise detrapping curves. We then fit a sum of exponentials with fixed time constants $\tau_i = 1,10,100,1000,10000,100000$ seconds to the detrapping curves. 

\begin{equation}
Q(t)=\displaystyle\sum_{i=1}^6\bigg[N_i
.\bigg(1-\exp\big({\frac{-t}{\tau_i}\big)}\bigg)\bigg] \\
\label{eq:SigmaTau}
\end{equation}

The $N_i$ coefficient of each exponential term gives the number of traps in that detrapping time constant bin. The result of this fit is a time constant spectrum, a vector $N$. For most, but not all,  detectors, this time constant spectrum $N$ is very similar in shape across the detector (exceptions will be explained in section \ref{sec:results}). 

We then normalize this vector $N$ to a total of 1 such that it represents the fraction of traps in each time constant bin. We fix this vector and go back and fit every pixel on the detector for a total number of traps. The result of this analysis is a single vector with 5 values corresponding to the fraction of traps in each time constant bin, and a 4096x4096 pixel total trap map corresponding to the maximum persistence signal in each pixel. This can then be divided by the illumination frame to get a trap density map (trapped charge as a percentage of illumination). This dataset can then be used by the pipeline for persistence prediction and correction, described in section \ref{sec:pipelincorrection}. 

Note that this approach requires zero knowledge of the underlying physics causing the persistence. This is a limitation of the method, but also an asset. As discussed in section \ref{sec:results} there are likely several different causes of persistence in the range of detectors we've tested, and this method allows us to robustly fit (and correct for) the persistence from all of them. One can view this spectrum as similar to the spectra obtained from Deep Level Transient Spectroscopy (DLTS), but sensitive to much longer time constants.\cite{rubaldo14} Some physics can be teased out of this fitting method, as was done in Ives et al. 2020, \cite{ives20} in which the movement of the peak in the detrapping spectrum as a function of temperature could be fit to a Shockley-Read-Hall (SRH) detrapping model giving an estimated trap energy. 

\section{RESULTS}
\label{sec:results}

\begin{figure} [htbp]
   \begin{center}
   \begin{tabular}{c}  
   \includegraphics[width=17cm]{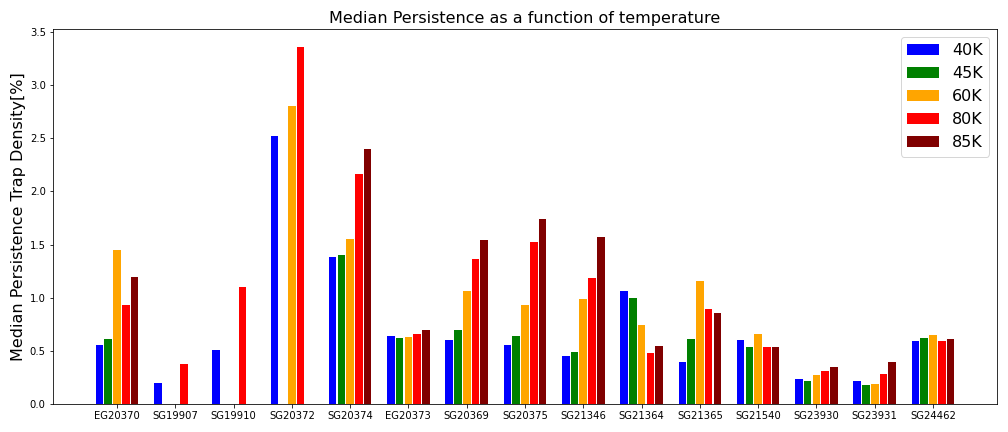}
   \end{tabular}
   \end{center}
   \caption[example] 
   {\label{fig:medianpersistencevstemp} Median persistence trap density for each detector at the tempeartures tested. Details of each detector's persistence behaviour are listed in table \ref{tab:persistenceresults}.}
   \end{figure}

Table \ref{tab:persistenceresults} gives the results of the median trap density of each detector at 40 and 80 K, but also notes detectors where the median trap density does not tell the whole story (e.g. high contact resistance devices, devices with high stress, etc.). Figure \ref{fig:medianpersistencevstemp} gives the median persistence as a function of temperature for each detector. Figures \ref{fig:trapdensitymaps40K} and \ref{fig:trapdensitymaps80K} in \ref{sec:appendixA} show the trap density maps at 40 and 80 K for all detectors.

\begin{table}[p]
\caption{Persistence results for the 15 tested detectors.} 
\label{tab:persistenceresults}
\begin{center}       
\begin{tabular}{|p{0.1\linewidth} | p{0.1\linewidth} |  p{0.15\linewidth}|p{0.5\linewidth}|}
\hline
\rule[-1ex]{0pt}{3.5ex}  Detector & Median Trap Density (40 K, 80 K) & Type & Notes \\
\hline
\rule[-1ex]{0pt}{3.5ex}  EG20370 &0.6\%, 0.9\% & 65K peak & Extensively tested over many years and cooldowns as it was the first detector.\cite{george21}\\
\hline 
\rule[-1ex]{0pt}{3.5ex}  SG19907 & 0.2\%, 0.4\%  & increasing or 65K peak & Lowest persistence detector. No 60K data collected, so can either be a increasing or 65K peak detector \\
\hline 
\rule[-1ex]{0pt}{3.5ex}  SG19910 & 0.5\%, 1.1\%& increasing or 65K peak & no 60K data collected, so can either be a increasing or 65K peak detector\\
\hline 
\rule[-1ex]{0pt}{3.5ex}  SG20372 &2.5\%, 3.6\% & increasing & This was the highest persistence detector with a very wide distribution of trap densities, with some pixels showing up to 6\% trap density.\\
\hline 
\rule[-1ex]{0pt}{3.5ex}  SG20374 & 1.4\%, 2.2\% & increasing & Very hot pixel regions begin to show negative signal at temperatures above 80 K.\\
\hline 
\rule[-1ex]{0pt}{3.5ex}  EG20373 &0.6\%, 0.7\% & flat & relatively flat persistence versus temperature, at 80 K and above hot pixel regions begin to have additional long-time constant traps.\\
\hline 
\rule[-1ex]{0pt}{3.5ex}  SG20369 &0.6\%, 1.5\% &increasing & Some long time constant traps at high temperatures that decrease as temperatures drop.\\
\hline 
\rule[-1ex]{0pt}{3.5ex}  SG20375 &0.6\%, 1.7\% &increasing & Consistent time constant spectra with temperature, with only decreases in magnitude.\\
\hline
\rule[-1ex]{0pt}{3.5ex}  SG21346 &0.5\%, 1.6\% &increasing & Very hot pixel regions begin to show negative signal at temperatures above 80 K.\\
\hline
\rule[-1ex]{0pt}{3.5ex}  SG21364 &1.1\%, 0.5\% &decreasing  & high persistence especially in the middle of the array at low temperatures. This detector also showed a significant drop in Js QE in the center of the array and strong crosshatch, as well as hot pixel regions with long-time constant traps at 80 and 85 K.
 \\
\hline 
\rule[-1ex]{0pt}{3.5ex}  SG21365 &0.4\%, 0.9\% &65K peak & The median trap density at 60K was 1.2\%\\
\hline  
\rule[-1ex]{0pt}{3.5ex}  SG21540 &0.6\%, 0.5\% &high contact resistance & median trap density in good regions stays relatively flat, but high contact resistance regions have much higher persistence at low temperatures so there is a long tail on the trap density histogram. \\
\hline 
\rule[-1ex]{0pt}{3.5ex}  SG23930 &0.2\%, 0.4\% &high contact resistance & median trap density in good regions actually increases with temperature, but high contact resistance regions have much higher persistence at low temperatures so there is a long tail on the trap density histogram that is much more prominent at low temperatures. \\
\hline 
\rule[-1ex]{0pt}{3.5ex}  SG23931 &0.2\%, 0.4\% &high contact resistance & median trap density in good regions actually increases with temperature, but high contact resistance regions have much higher persistence at low temperatures so there is a long tail on the trap density histogram that is much more prominent at low temperatures. \\
\hline 
\rule[-1ex]{0pt}{3.5ex}  SG24462 &0.6\%, 0.6\% &flat & Most of the array is flat, but edges of the array and upper left corner show increasing persistence with temperature and increasing long time constant traps.\\
\hline 
\end{tabular}
\end{center}
\end{table}

 Most of the detectors could be fit with a single spectrum of detrapping time constants with only the number of traps varying across the detector. However, in a few detectors this was not true. We categorized the detectors based on the behaviour of the median trap density with temperature. Of the 15 detectors tested, we could categorize 13 of them, with the remaining 2 unable to be categorized based on data only being available at a subset of test temperatures.

\subsection{Increasing detectors}
\label{sec:increasing}
The largest number of detectors (five, up to seven) have detrapping time constant spectra that are very similar across the whole detector, with only a spatial variation in numbers of traps. At least five detectors have a monotonically increasing trap density with temperature (roughly doubling between 40 and 85 K), allowing a free selection of operating temperature anywhere between 40-85 K. This allows for optimization of other detector parameters over the full temperature range. Typically, areas of these detectors with increased hot pixels also have increased persistence, but not always. 

\begin{figure} [htbp]
   \begin{center}
   \begin{tabular}{c}  
   \includegraphics[width=16cm, trim=1.5cm 3.5cm 0.0cm 3cm, clip]{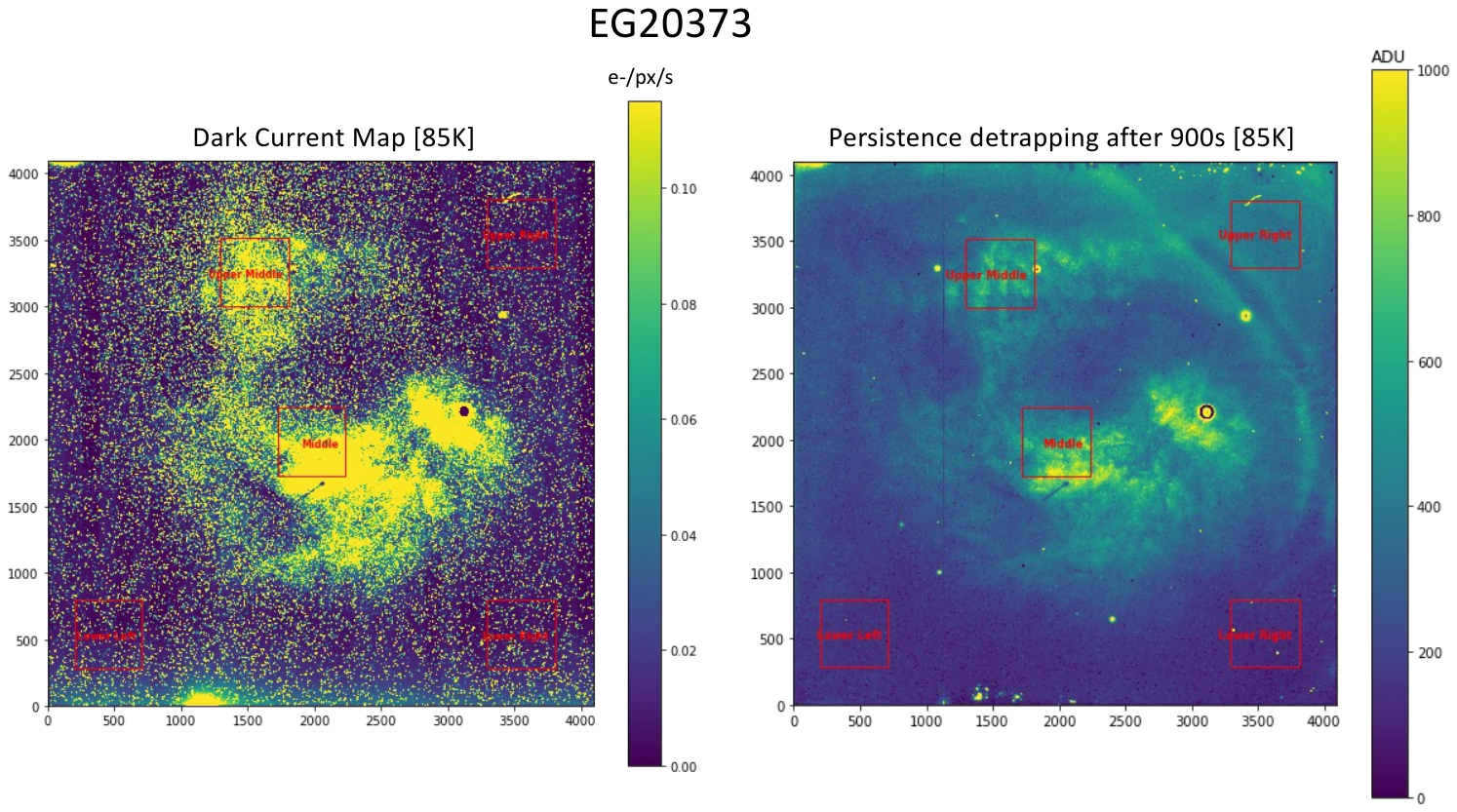}\\
      \includegraphics[width=17cm, trim=1.0cm 3.5cm 2.0cm 4cm, clip]{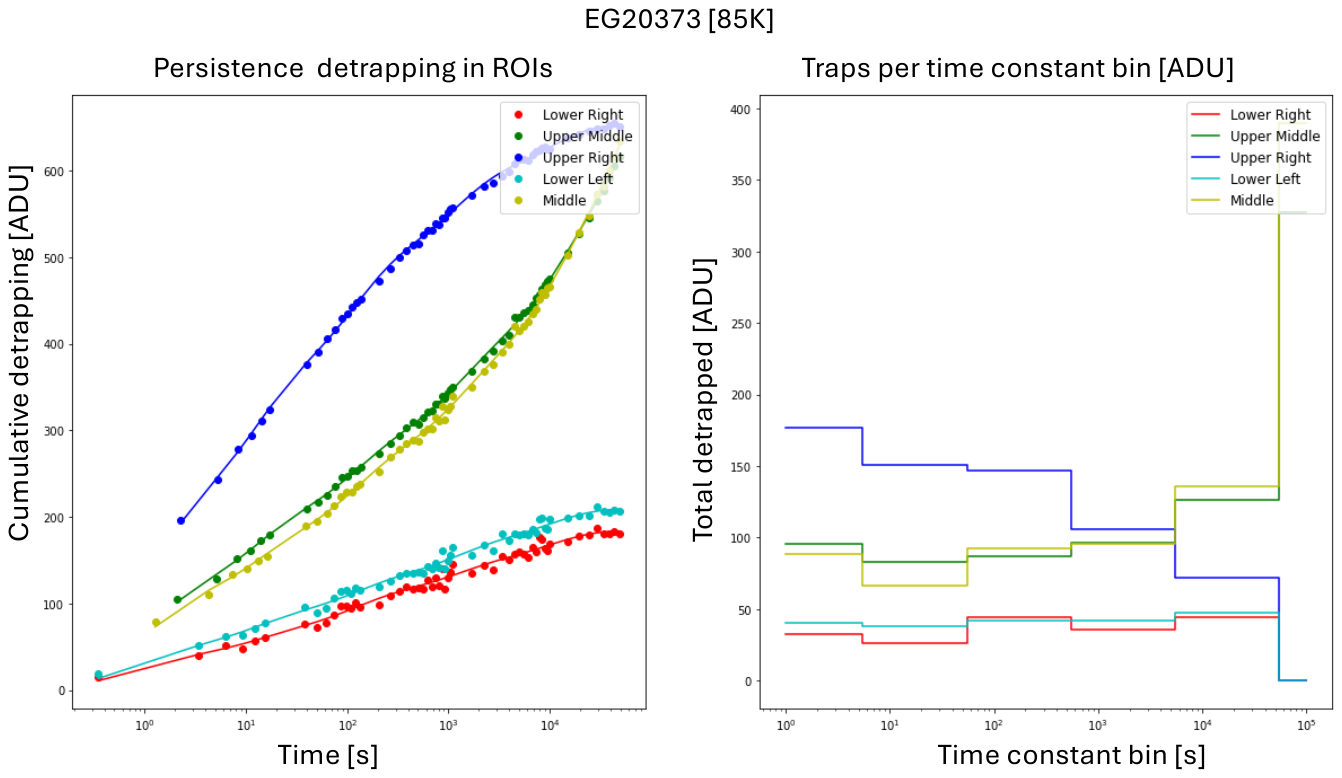}
   \end{tabular}
   \end{center}
   \caption[example] 
   { \label{fig:EG20373-85K-darkpersistence} Top Left: Dark current map of the EG20373 detector at 85 K. Top Right: Persistence detrapping after about 900s. The five labeled regions of interest correspond to the detrapping curves shown in the plots below. Bottom Left: Detrapping curves and fits to the data in the five regions of interest shown above. Bottom Right: Time constant spectra for the five regions of interest. Note that the regions with many hot pixels [Middle and Upper middle] show excess long time constant traps. At temperatures 60K and below, this behaviour in the hot pixel regions is no longer seen and these pixels have similar detrapping curves to the rest of the array. This detector showed a flat median persistence as a function of temperature, but the hot pixels regions are clearly an exception.}
   \end{figure}

A few detectors at 80 and 85 K showed hot pixel regions with excess long-time constant traps, while others with very hot pixel regions began to show negative signal at long times at temperatures above 80 K. This is associated with regions of high crosshatch pattern in some detectors. These detectors are challenging to fit with our model at high temperatures, which assumes that the time constant spectrum is the same over the whole detector. As individual pixel fits is difficult with the measurements we obtain, a potential solution would be to identify all high dark current pixels using dark current maps, and then fit the two populations of pixels separately, producing two time constant spectra and two trap maps, each covering one population of pixels. Figure \ref{fig:EG20373-85K-darkpersistence} shows an example of one detector that has this long time constant behaviour in hot pixel regions at 80 and 85 K. In detectors with these high dark current/high persistence regions, it is advisable to choose a lower operation temperature to avoid the excess long-time constant traps.

\subsection{Flat detectors}
\label{sec:flat}
Two detectors show flat persistence with temperature. These detectors otherwise behave similarly to the increasing detectors. 

\subsection{Decreasing detectors}
\label{sec:decreasing}
One detector (SG21364) showed decreasing persistence with temperature. At low temperatures, especially the center of the array showed very high persistence. This detector also showed a significant drop in Js QE in the center of the array and strong crosshatch pattern. For further discussion of the crosshatch pattern, see Bezawada et al. 2026 (these proceedings).\cite{bezawada26} Also to be noted is that in the hot pixel regions of this detector, there are similar long-time constant traps as seen in the increasing and flat detectors.

\subsection{65K persistence peak}
\label{sec:persistencehump}
\begin{figure} [htbp]
   \begin{center}
   \begin{tabular}{c}  
   \includegraphics[width=17cm, trim=0cm 0cm 0cm 0cm, clip]{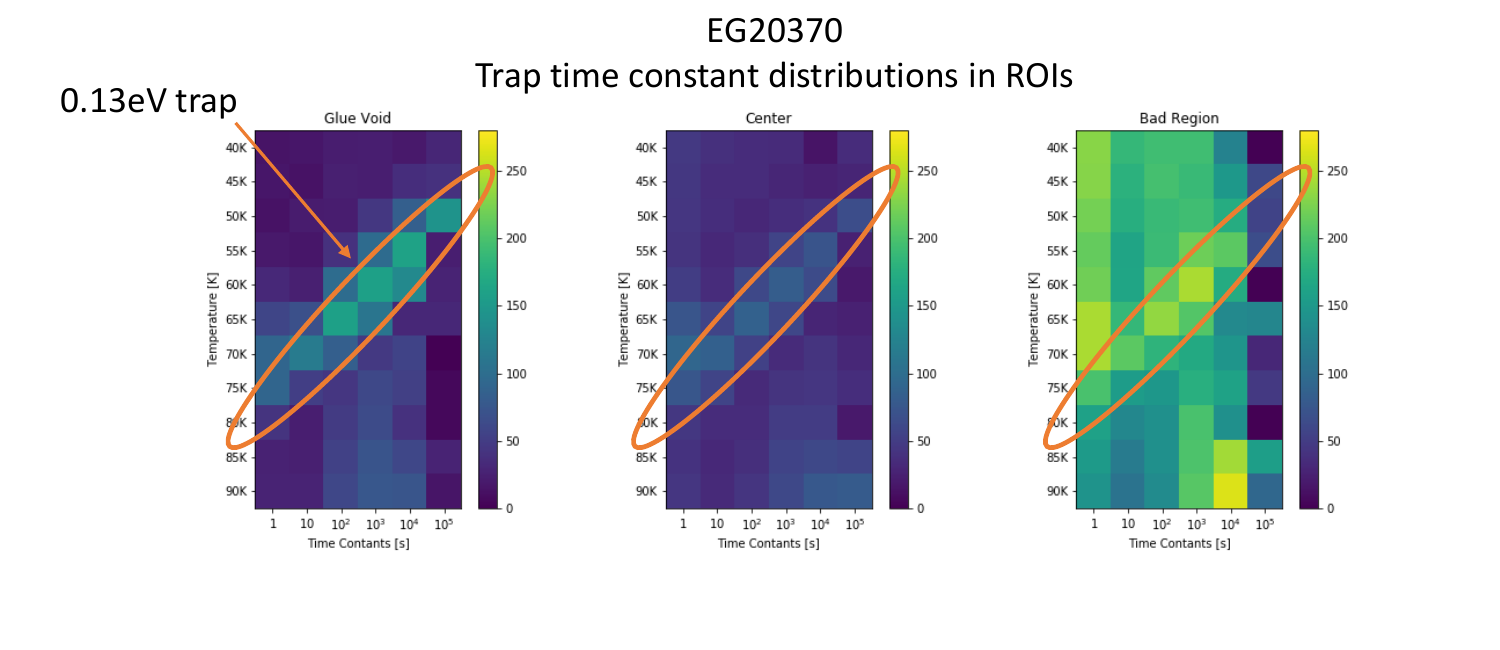}
   \end{tabular}
   \end{center}
   \caption[example] 
   { \label{fig:EG20370waterfall} Detector EG20370 waterfall diagram reproduced from Ives et. al 2020\cite{ives20}. The peak in the detrapping time constant spectra moves as a function of temperature, indicating a trap with an energy around 0.13eV. In particular, detrapping peaks in the range of 100s to 1000s of seconds at intermediate temperatures, which impacts science exposures most dramatically.}
   \end{figure}

 These detectors similarly show time constant spectra that are similar across the whole detector with only the trap density spatially varying. However, instead of the average trap density monotonically increasing, decreasing, or remaining flat with temperature, these detectors show an increase in persistence between approximately 50-75 K with the peak in the time constant spectrum varying with temperature (moving to longer times at lower temperature) which we attribute to a trap with an energy around 0.13 eV (see figure \ref{fig:EG20370waterfall}). For more details, see Ives et al. 2020.\cite{ives20} A very important characteristic of this type of persistence is not only is there an overall increase in persistence at intermediate temperatures, but that the time constants of the persistence detrapping fall right in the range where they maximally affect subsequent exposures (100s-1000s of seconds) around this persistence peak. 

 As the very first detector tested (EG20370) showed this behaviour, we made the recommendation for instruments to choose an operating temperature below 45K or above 75K, but not in between. Unfortunately, it is not possible to know before testing an individual detector whether it will show this persistence peak at intermediate temperatures, and usually instrument cryogenic design and operation temperatures are decided well in advance of detector procurement and testing. However, of the 15 detectors tested only 2 detectors definitively showed this behavior, which means that this recommendation should now be reconsidered.

\subsection{High contact resistance pixels}
\label{sec:contactresistance}
In the three detectors tested that show high contact resistance pixels, the persistence behavior varies spatially over the detector. Pixels with high contact resistance show increasing persistence with lower temperatures, while the pixels on the detector without the contact resistance issue show similar behavior to the flat or increasing detectors. Additionally, the high contact resistance pixels detrap all of their persistence very rapidly (within the first several hundred seconds) which is very different than the persistence seen in other types detectors and suggests a different mechanism (such as RC discharge) or a different population of traps. 

   \begin{figure} [htbp]
   \begin{center}
   \begin{tabular}{c}  
   \includegraphics[width=17cm, trim=0cm 1.5cm 0cm 0cm, clip]{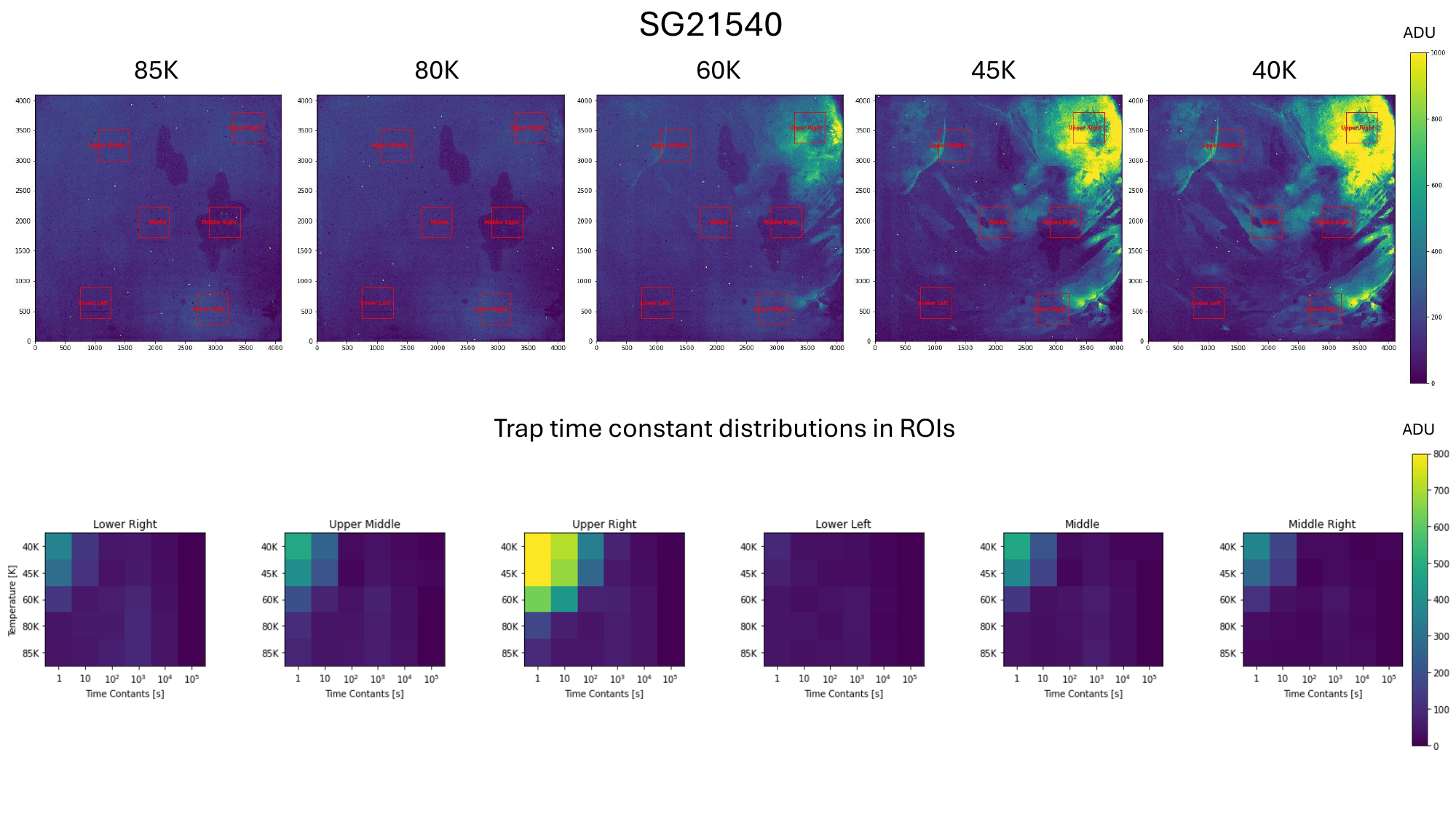}
   \end{tabular}
   \end{center}
   \caption[example] 
   { \label{fig:highcontact-resistance} Detector SG21540, which is heavily affected at low temperatures by high contact resistance pixels. Top row: Persistence signal detrapped after approximately 900s during measurement at different temperatures with some Regions of Interest (ROIs) highlighted. Bottom row: For each region of interest indicated, the time constant distribution from fits of the detrapping curves in that ROI, presented as a waterfall diagram. Note at temperatures of 60K and below, the quantity of short time constant traps in the fit increases rapidly.}
   \end{figure} 

These detectors are challenging to fit with our model at low temperatures, which assumes that the time constant spectrum is the same over the whole detector. Similar to the case of the high dark current pixels, a potential solution here would be to identify all high contact resistance pixels using noise maps, and then fit the two populations of pixels separately, producing two time constant spectra and two trap maps, each covering one population of pixels. 

\section{MODELING}
\label{sec:modeling}
For every detector tested, we have a time constant spectra, a maximum trap map, and a persistence density map. This can be used as input into a simulation of detector exposures. For example, using an input source image, we can simulate the buildup of persistence charge in the traps over an image. Then upon resetting the detector, simulate the decay of the persistence charge which will be collected in following exposures. Practically this is done by creating an image array for each time constant bin and then keeping track of the trapped and detrapped charges over the exposure time. Tulloch and George\cite{tulloch19} and Padovani et al. 2022\cite{padovani22} both show examples of this. 

\begin{figure} [ht]
   \begin{center}
   \begin{tabular}{c}  
   \includegraphics[height=3cm]{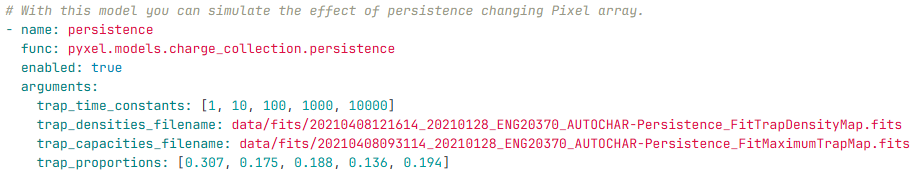}
   \end{tabular}
   \end{center}
   \caption[example] 
   { \label{fig:persistenceyaml} An example of the yaml configuration file to run the persistence model in pyxel.}
   \end{figure}

We have implemented this persistence model in the pyxel detector simulation framework. The pyxel detector frameworks allows the user to create a detector object with different attributes, and then run different models during an exposure to simulate different detector effects (for example, dark current generation or readout noise generation).\cite{lemmel26} The inputs required for the persistence model are a maximum trap map, a trap density map, and a time constant ratio vector that corresponds to the fraction of traps in each time constant bin. The user is free to define their own time constant bins and corresponding trap ratio vectors in the configuration file, but ideally this input data is generated by a characterization of a real detector. Figure \ref{fig:persistenceyaml} shows an example of the configuration. This example (including the input trap map fits files measured from the EG20370 detector) are available as an example notebook in the pyxel-data repository for readers who would like to try this out for themselves. \url{https://gitlab.com/esa/pyxel-data/-/blob/master/examples/exposure/exposure_persistence-H4RG.ipynb} 

\section{PIPELINE CORRECTION}
\label{sec:pipelincorrection}
Image persistence in NIR detectors is a long-standing issue and is seen in many of ESO's NIR instruments in cases where bright sources or calibrations are observed. One of the goals of the persistence characterization campaign was to produce characterization data that could be used in a persistence correction algorithm. Full persistence characterization data at the instrument operating temperatures has now been collected for the ERIS H2RG detectors (SPIFFIER and NIX) as well as the MOONS and MICADO H4RG detectors. Similar characterization campaigns are planned for future instruments.

We have developed an algorithm based on the model described here where the traps are tracked throughout the exposures leading up to a science image. A cumulative persistence decay map can then be calculated for each science frame. Data taken with ESO's telescopes is typically proprietary for 1 year, so these persistence maps need to be calculated by the ESO Paranal Science Operations group. The goal is that these persistence maps can be put into the ESO archive as calibration frames associated with each science exposure. In the rare case where a science exposure is contaminated by persistence, it will be left to the user to subtract the persistence calibration frame. Padovani et al. 2022\cite{padovani22} shows an example of the persistence correction algorithm applied to the persistence caused by a bright ERIS/SPIFFIER wavelength calibration frame.

   \begin{figure} [ht]
   \begin{center}
   \begin{tabular}{c} 
   \includegraphics[height=7cm]{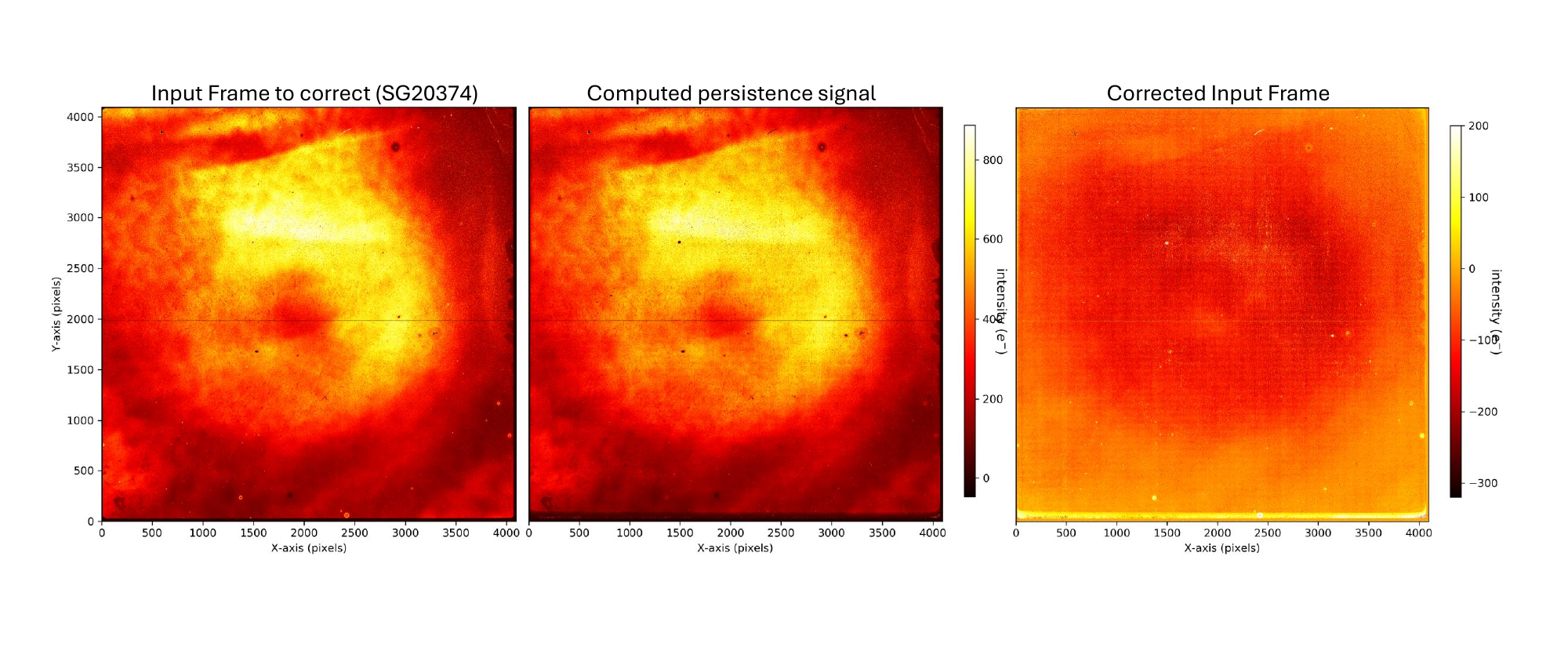}
   \end{tabular}
   \end{center}
   \caption
   {\label{fig:Correction_20374} Left: A 300s dark input frame taken after a saturating flat field exposure in detector 20374, heavily contaminated with persistence. Middle: Persistence signal computed with the correction algorithm. Right: Corrected input frame. Note a slight over-correction in the middle of the frame as well as an under-correction in the strip along the bottom where the mask above the detector cast a shadow during the test campaign in which we collected the data to create the persistence trap maps. Credit: Mark Neeser}
   \end{figure} 

As has been reported previously\cite{tulloch18, tulloch19}, the gain of the persistence signal is less than that of photoelectrons or dark current electrons. This means that subtracting off persistence signal should introduce relatively less noise than, for example, subtracting off the same magnitude of sky signal or dark current. There is therefore little downside to subtracting off persistence signal, with the caveat that the prediction algorithm is not perfect and may slightly under- or over-correct. However, this is superior to leaving uncorrected persistence signal in the science image. 

One important factor in this approach is that the persistence trap maps and time constant vectors must remain stable over time for this strategy to work, as the persistence trap map measurement is not a calibration that can be regularly performed on the telescope. Detectors typically stay in instruments for their entire lifetimes, which can exceed a decade of service. In lab testing the maps have remained stable from cool-down to cool-down (the EG20370 detector was tested in 11 cool-downs over a 4 year time span). However, at the telescope, the initially measured large persistence trap amplitude in the ERIS/NIX detector decreased with time, until at this point persistence is almost never seen in the ERIS/NIX detector, and therefore the persistence correction pipeline is practically not needed with this instrument. We do not have an explanation for this, but speculate it could be due to either a slightly different operation temperature at the telescope or perhaps an effect of the detector staying continuously cold for months or years at a time. MOONS is now in commissioning and we will soon have an additional 4 detectors on sky, with which to test this correction strategy.

\section{CONCLUSION}
\label{sec:conclusion}

In this study spanning from 2019-2025, fifteen H4RG-15 detectors for MOONS and MICADO were characterized for persistence detrapping between 40-85 K. The detectors were categorized into 5 types based on the behaviour of the median persistence trap density versus temperatures: increasing, flat, decreasing, 65K peak, and high-contact resistance. For most detectors, a model with a single time constant spectra plus a trap density/maximum trap map is sufficient to describe the persistence behavior over the whole array at a single temperature. In some detectors, regions of high dark current show long-time constant traps appearing above 80 K. In these detectors at high temperatures, one time constant spectra may not be sufficient to describe persistence behavior in hot-pixel regions and different populations of pixels may need to be categorized and fit separately using dark current maps. For high-contact resistance detectors, the pixels suffering from high contact resistance show anomalous high persistence at low temperatures that detraps very quickly, over 100s of seconds. More than one time constant spectra may be required to fit the data from these detectors and different populations of pixels may need to be categorized and fit separately using noise maps. 

Only two of fifteen detectors definitively showed a peak in persistence at 65K. This means that our previous recommendation of choosing an instrument operation temperature below 45K or above 75K to avoid the peak is not valid for the majority of detectors. Operation temperatures may be chosen freely for the majority of detectors, in tradeoff with other performance parameters such as noise (generally lower at lower temperatures) and QE (generally higher at higher temperatures)\cite{bezawada26}. Only after full detector characterization at temperatures from 40-85 K and a tradeoff analysis of performance requirements based on science goals can the optimal operation temperature be chosen. Every detector is unique, and every one will likely have an optimal operation temperature and an acceptable range. \textbf{We therefore recommend implementing a flexible thermal interface in the instrument design phase allowing for the detector operating temperature to be chosen after the detectors have been fully characterized.}

Based on this characterization data, a persistence prediction and correction model was developed. This model is implemented in pyxel, and an example notebook using the characterization data from EG20370 is available for the community to try out. The correction algorithm is also implemented in the ESO High level Data Reduction Library (HDRL)\cite{gabasch20}, which allows automatic creation of persistence correction frames based on the detector exposures in the hours preceding the potentially contaminated science exposure. This has already been deployed for the ERIS NIX/SPIFFIER instruments, and we have the data to also implement this for MOONS and MICADO. 

We plan to collect persistence characterization data also for future ESO instruments to create trap maps and time constant spectra. As we learn more about the different types of persistence seen in different detectors, the model may need to be optimized to take into account different pixel populations. This is the subject of future work.

\appendix    
\section{Test conditions}
\label{sec:appendixA}
Here we include the test facility and test temperatures at which datasets were obtained for each detector. Table \ref{tab:detectors} includes these details.

\begin{table}[ht]
\caption{Detectors tested in this campaign and the temperatures at which persistence data was obtained.} 
\label{tab:detectors}
\begin{center}       
\begin{tabular}{|l|l|l|l|} 
\hline
\rule[-1ex]{0pt}{3.5ex}  Detector & Instrument & Test facility & Temperature [K] \\
\hline
\rule[-1ex]{0pt}{3.5ex}  EG20370 & MOONS & CRISLER  & 40, 45, 50, 55, 60, 65, 70, 75, 80, 85, 90 \\
\hline 
\rule[-1ex]{0pt}{3.5ex}  SG19907 & MOONS & CRISLER  & 41, 80 \\
\hline 
\rule[-1ex]{0pt}{3.5ex}  SG19910 & MOONS & CRISLER  & 40, 80 \\
\hline 
\rule[-1ex]{0pt}{3.5ex}  SG20372 & MOONS & CRISLER  & 40, 60, 80 \\
\hline 
\rule[-1ex]{0pt}{3.5ex}  SG20374 & MOONS & CRISLER  & 40, 45, 60, 80, 85 \\
\hline 
\rule[-1ex]{0pt}{3.5ex}  EG20373 & MICADO/HARMONI & FIAT & 40, 45, 60, 80, 85  \\
\hline 
\rule[-1ex]{0pt}{3.5ex}  SG20369 & MICADO & FIAT  & 40, 45, 60, 80, 85 \\
\hline 
\rule[-1ex]{0pt}{3.5ex}  SG20375 & MICADO & FIAT  & 40, 45, 60, 80, 85 \\
\hline
\rule[-1ex]{0pt}{3.5ex}  SG21346 & MICADO & FIAT  & 40, 45, 60, 80, 85\\
\hline
\rule[-1ex]{0pt}{3.5ex}  SG21364 & MICADO & FIAT  & 40, 45, 60, 80, 85\\
\hline 
\rule[-1ex]{0pt}{3.5ex}  SG21365 & MICADO & FIAT  & 40, 45, 60, 80, 85 \\
\hline  
\rule[-1ex]{0pt}{3.5ex}  SG21540 & MICADO & FIAT  & 40, 45, 60, 80, 85 \\
\hline 
\rule[-1ex]{0pt}{3.5ex}  SG23930 & MICADO & FIAT  & 40, 45, 60, 80, 85 \\
\hline 
\rule[-1ex]{0pt}{3.5ex}  SG23931 & MICADO & FIAT  & 40, 45, 60, 80, 85\\
\hline 
\rule[-1ex]{0pt}{3.5ex}  SG24462 & MICADO & FIAT  & 40, 45, 60, 80, 85\\
\hline 
\end{tabular}
\end{center}
\end{table}

\begin{figure} [htbp]
   \begin{center}
   \begin{tabular}{c}  
   \includegraphics[width=17cm,trim=1.5cm 3.0cm 0cm 1cm, clip]{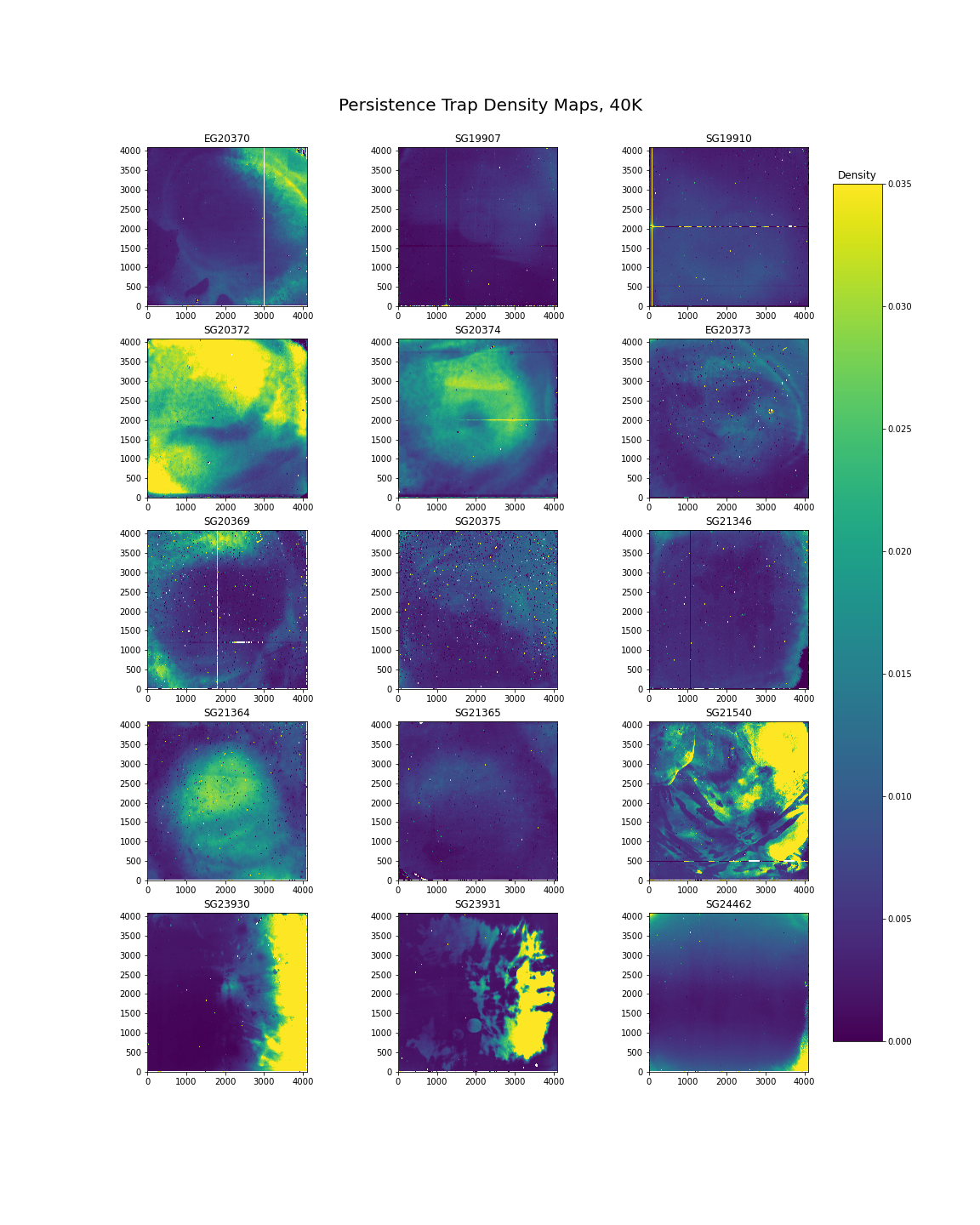}
   \end{tabular}
   \end{center}
   \caption[example] 
   {\label{fig:trapdensitymaps40K} Trap Density maps for all detectors at 40 K.}
\end{figure}

\begin{figure} [htbp]
   \begin{center}
   \begin{tabular}{c}  
   \includegraphics[width=17cm,trim=1.5cm 3.0cm 0cm 1cm, clip]{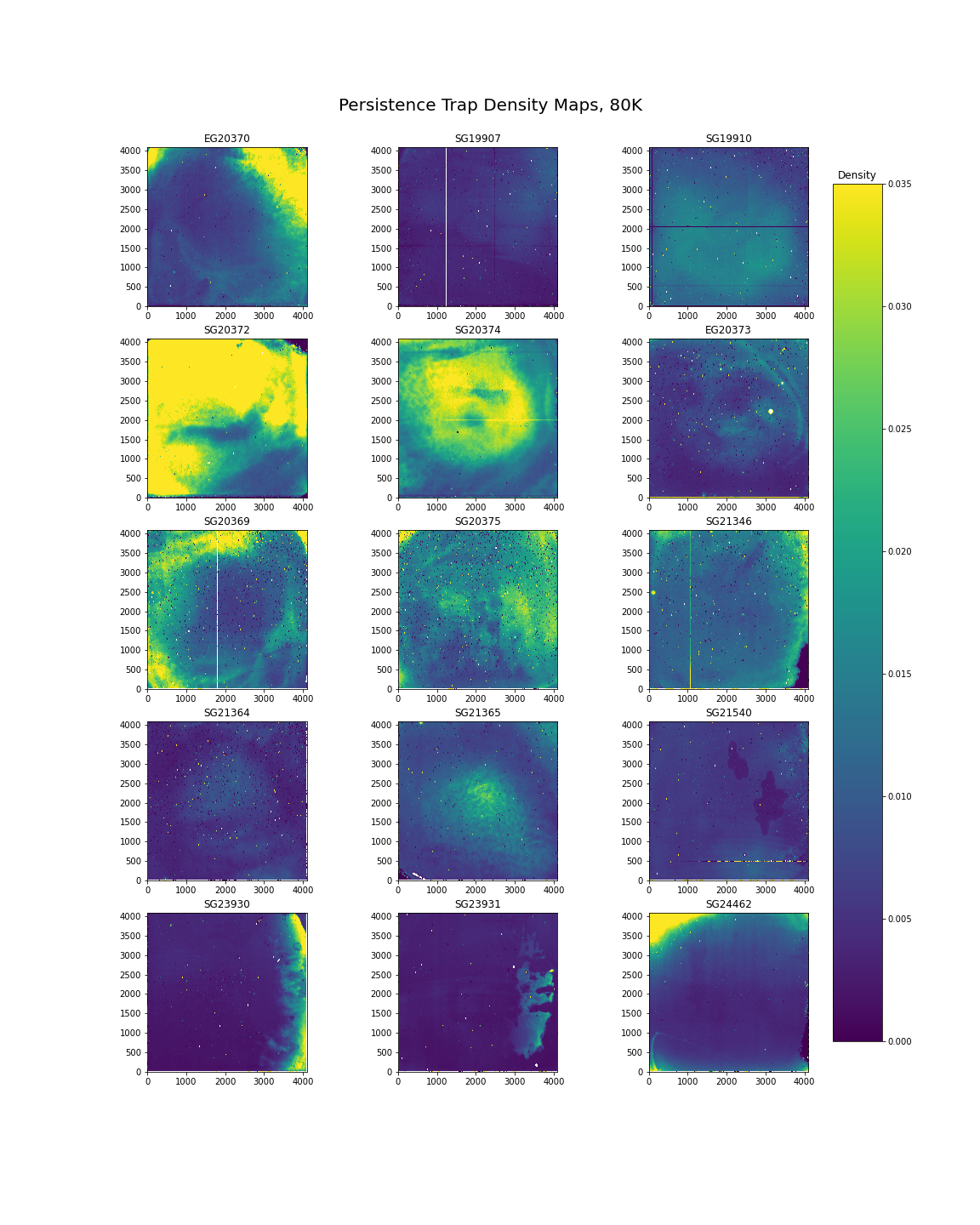}
   \end{tabular}
   \end{center}
   \caption[example] 
   {\label{fig:trapdensitymaps80K} Trap Density maps for all detectors at 80 K.}
\end{figure}

Figures \ref{fig:trapdensitymaps40K} and \ref{fig:trapdensitymaps80K} show the trap density maps at 40 and 80 K. These temperatures were chosen as the most detectors had available data at these temperatures and the 80 and 85 K maps are quite similar in most detectors. Note that trap density is defined as the fraction of charge carriers that can be trapped at all time constants. 

\section* {Code, Data, and Materials Availability} 
The H4RG detector data used in this paper was acquired and analyzed in the ESO detector labs and could be made available upon reasonable request to the corresponding author. The pyxel detector modeling framework is available on ESA's gitlab: \url{https://gitlab.com/esa/pyxel}. 

\newpage  

\bibliography{report} 
\bibliographystyle{spiebib} 

\end{document}